\documentclass[12pt,aps,pra,groupedaddress,amssymb,amsfonts,nofootinbib,tightenlines]{revtex4} 
\usepackage{graphicx} 

\graphicspath{{./}{}}
\usepackage{times,color}
\definecolor{purple}{rgb}{0.625,0.125,0.9375}

\usepackage{ekqc_b}

\ignore{
rm -R /tmp/state; mkdir /tmp/state
cp `texfls state.log | perl -e '$a=<>; $a =~ s:/\S*(revtex4|natbib)\S*(\s|$)::g; print "$a\n"'` /tmp/state
find /tmp/state -type f -name '*.pdf' -exec pdftops -eps {} \; -exec rm {} \;
find /tmp/state -type f -name '*.jpg' -exec perl -e '$f=shift; $g=f$; $g=~s/.jpg/.eps/; exec("convert $f $g");' {} \; -exec rm {} \;
perl -i -n -e 's/^

cp /usr/share/texmf/tex/latex/tools/calc.sty /tmp/state/phcalc.sty
perl -i -n -e 's/ProvidesPackage\{calc\}/ProvidesPackage\{phcalc\}/; print;' /tmp/state/phcalc.sty
perl -i -n -e 's/usepackage\{calc\}/usepackage\{phcalc\}/; print;' /tmp/state/ekqc_b.sty

(cd /tmp/state; tar czvf state.tar.gz *)
}

\ignore{
rm -R /tmp/state; mkdir /tmp/state
cp `texfls state.log | perl -e '$a=<>; $a =~ s:/\S*(revtex4|natbib)\S*(\s|$)::g; print "$a\n"'` /tmp/state
find /tmp/state -type f -name '*.pdf' -exec pdftops -eps {} \; -exec rm {} \;
find /tmp/state -type f -name '*.jpg' -exec perl -e '$f=shift; $g=f$; $g=~s/.jpg/.eps/; exec("convert $f $g");' {} \; -exec rm {} \;
perl -i -n -e 's/^
perl -i -n -e 's/graphics\///; print;' /tmp/state/state.tex

cp /usr/share/texmf/tex/latex/tools/calc.sty /tmp/state/phcalc.sty
perl -i -n -e 's/ProvidesPackage\{calc\}/ProvidesPackage\{phcalc\}/; print;' /tmp/state/phcalc.sty
perl -i -n -e 's/usepackage\{calc\}/usepackage\{phcalc\}/; print;' /tmp/state/ekqc_b.sty

cp /usr/share/texmf/tex/latex/base/ifthen.sty /tmp/state/phifthen.sty
perl -i -n -e 's/ProvidesPackage\{ifthen\}/ProvidesPackage\{phifthen\}/; print;' /tmp/state/phifthen.sty
perl -i -n -e 's/usepackage\{ifthen\}/usepackage\{phifthen\}/; print;' /tmp/state/ekqc_b.sty

(cd /tmp/state; tar czvf state.tar.gz *)
}

\begin{document}
\title{Fault-Tolerant Postselected Quantum Computation: Schemes}
\author{E. Knill}
\email[]{knill@boulder.nist.gov}
\affiliation{Mathematical and Computational Sciences Division,
             National Institute of Standards and Technology,
             Boulder, CO 80305} 

\date{\today}

\begin{abstract}
Postselected quantum computation is distinguished from regular quantum
computation by accepting the output only if measurement outcomes
satisfy predetermined conditions.  The output must be accepted with
nonzero probability.  Methods for implementing \mbox{postselected} quantum
computation with noisy gates are proposed. These methods are based on
error-detecting codes. Conditionally on detecting no errors, it is
expected that the encoded computation can be made to be arbitrarily
accurate. Although the probability of success of the encoded computation
decreases dramatically with accuracy, it is possible to apply the
proposed methods to the problem of preparing arbitrary stabilizer
states in large error-correcting codes with local residual
errors. Together with teleported error-correction, this may improve
the error tolerance of non-postselected quantum computation.
\end{abstract}


\maketitle

\section{Introduction}
\label{sect:introduction}

An important problem in the theory of scalable quantum computation is
to improve the maximum gate error rate for quantum gates at which it
is possible to quantum compute fault
tolerantly. In~\cite{knill:qc2003b}, it was shown that if the errors
are all detected, then it is possible to quantum compute with up to
$50\;\%$ error probability per Bell measurement. The techniques used
in\cite{knill:qc2003b} show that for depolarizing errors that are not
detected, high error rates can be tolerated if certain entangled
states can be prepared with sufficiently small, effectively
independent errors. To prepare the required states, it suffices to use
postselected quantum computation. Postselected quantum computation has
the property that computations are accepted only if predetermined
conditions are satisfied. The conditions must be satisfied with
nonzero probability. Here, methods are proposed for implementing
postselected quantum computation in the presence of noisy gates. The
basic idea is to use a simple error-detecting code to encode a
computation so that errors at a sufficiently low rate are detected. If
the output is accepted only if no error is detected (that is, if the
output is \emph{postselected}), then the conditional probability of error
can be reduced arbitrarily.  The states required for
(non-postselected) fault-tolerant quantum computation can then be
prepared in an encoded form with some probability of success. To use
them, the underlying error-detecting code is decoded with further
error-detection to eliminate remaining errors. If the error-detecting
code is obtained by concatenation, the final state can be obtained so
that residual error is local with error-rate determined by the error
model and the complexity of the decoding process.

The description of the methods in the next sections assumes the
notation and knowledge of the teleportation techniques and
stabilizer codes discussed in~\cite{knill:qc2003b}. The basic ideas
presented here are based on the large body of work on fault-tolerant
methods using stabilizer codes that have been developed by the quantum
information science
community~\cite{shor:qc1996a,kitaev:qc1996a,knill:qc1996b,aharonov:qc1996a,aharonov:qc1999a,knill:qc1997a,gottesman:qc1997a,knill:qc1998a,zalka:qc1996a,preskill:qc1998a,steane:qc1999a,gottesman:qc1999a,gottesman:qc2000b,knill:qc2000e,hughes:qc2002a,aharonov:qc2002a,duer:qc2002a,steane:qc2002a,steane:qc2002a,steane:qc2002b,steane:qc2003a}. For
further background information, see~\cite{knill:qc2003b} The main task
of this paper is to describe codes and networks for fault-tolerant
postselected quantum computation. That they may be expected to have
good fault-tolerance properties can be seen qualitatively. A detailed
analysis is still required to determine these properties
quantitatively and verify that they are as good as one would expect.

\section{Overview}
\label{sect:overview}

Fault-tolerant postselected quantum computation is easier to achieve
than regular fault-tolerant quantum computation because there is no
need to correct errors: it suffices to detect them. The disadvantage
is that the probabilities of success can be very small.  For
theoretical purposes, this is not a problem, provided that only states
of a bounded number of qubits need to be prepared.

To achieve fault-tolerant postselected quantum computation, a
four-qubit error-detecting code is combined with purification,
teleported error detection and transversal operations.  Transversal
operations involve applying gates only between pairs of corresponding
qubits in two blocks of four qubits used for encoding a state. By
concatenating this code, it is shown how to implement the accurate
operations needed to encode CSS stabilizer codes. With these
operations and the ability to encode an additional state with constant
encoded error, purification of this state leads to a universal set of
accurate gates and hence the full power of postselected quantum
computation. An interesting feature of the technique is that all
computations have finite depth. That postselected quantum computation
can be performed at constant depth has been pointed out in the context
of quantum complexity theory in~\cite{fenner:qc2003a}.

To obtain stabilizer states with bounded, local errors, postselected
encoded quantum computation is used to obtain an accurate stabilizer
state, encoded in the concatenated error-detecting code. The
concatenated error-detecting code is decoded bottom-up. If the state
is accepted only if no errors are detected during decoding, then the
output state is as desired.

\section{Basic Gates and Error Model}
\label{sect:basic gates}

The basic Clifford gate set used here consists of preparations of the
$+1$ eigenstates of $\sigma_z$ and $\sigma_x$, the controlled-not
(cnot) and $\sigma_z$ and $\sigma_x$ measurements. They suffice for
encoding and decoding CSS~\cite{calderbank:qc1996b} codes and states,
and are used for the primary concatenation. These gates, together with
the Hadamard gate, are referred to as the CSS gates. The Hadamard gate
is needed only at the top level of the concatenation hierarchy, where,
together with two prepared states, it is used to ensure
universality. Since the error-detecting code used permits a
transversal implementation of the Hadamard gate, it is possible to
add this gate without significantly affecting error-tolerance. To
enable all real quantum computations, preparation of the state
$\ket{\pi/8}=\cos(\pi/8)\ket{\biz}+\sin(\pi/8)\ket{\bio}$ (the $+1$
eigenstate of the Hadamard transform) is used. This is done fault
tolerantly by encoding it (with postselection) into the top level with
an encoded error related to the basic error rates.  Accurate encoded
cnot's and Hadamards are then used to purify the encoded
$\ket{\pi/8}$. Using the CSS gates and $\ket{\pi/8}$ preparation, it
is possible to implement quantum computation without complex phases.
Real-number quantum computation is equivalent in power to quantum
computation~\cite{bernstein:qc1997a}. But to complete the gate set,
one can add preparation of the state
$\ket{i\pi/4}={1\over\sqrt{2}}(\ket{\biz}+i\ket{\bio})$.  The network
symbols used for the gates and their definitions are shown in
Fig.~\ref{fig:gates}.

\begin{herefig}
\label{fig:gates}
\begin{picture}(0,3.8)(0,-3.3)
  \nputbox{0,.7}{t}{
     \setlength{\fboxrule}{0pt}\setlength{\fboxsep}{0pt}
     \fcolorbox[rgb]{.92,.92,.92}{.92,.92,.92}{\rule{6.2in}{0pt}\rule{0pt}{3.8in}}
  }
  \nputgr{-2.5,0}{c}{scale=.6}{prepZ}
     \nputbox{-2,0}{l}{$\ket{\biz}$}
  \nputgr{-2.5,-.5}{c}{scale=.6}{prepX}
     \nputbox{-2,-.5}{l}{${1\over\sqrt{2}}(\ket{\biz}+\ket{\bio})$}
  \nputgr{-2.5,-1}{c}{scale=.6}{prepY}
     \nputbox{-2,-1}{l}{$\cos(\pi/4)\ket{\biz}+i\sin(\pi/4)\ket{\bio}$}
 \nputgr{-2.5,-1.5}{c}{scale=.6}{prepH}
     \nputbox{-2,-1.5}{l}{$\cos(\pi/8)\ket{\biz}+\sin(\pi/8)\ket{\bio}$}

  \nputgr{.3,.0}{c}{scale=.6}{measz}
     \nputbox{.3,.2}{b}{Measure $\sigma_z$.}
  \nputgr{2.3,-1}{c}{scale=.6}{measz+}
     \nputbox{2.3,-.8}{b}{\begin{tabular}{l}
                           Measure $\sigma_z$,\\postselect on $+1$.
                         \end{tabular}}

  \nputgr{.3,-.8}{c}{scale=.6}{measx}
     \nputbox{.3,-.6}{b}{Measure $\sigma_x$.}
  \nputgr{2.3,-1.8}{c}{scale=.6}{measx+}
     \nputbox{2.3,-1.6}{b}{\begin{tabular}{l}
                             Measure $\sigma_x$,\\postselect on $+1$.
                          \end{tabular}}

  \nputgr{-2.5,-2.5}{c}{scale=.6}{had}
    \nputbox{-2.5,-2.3}{b}{${1\over\sqrt{2}}\qaop{1}{1}{1}{-1}$}
  \nputgr{0,-2.5}{c}{scale=.6}{cnot}
    \nputbox{0,-2.0}{b}{$\kets{\qbsym{ab}}{12}\rightarrow\kets{\qbsym{a(a{+}b)}}{12}$}
    \nputbox{-.3,-2.2}{b}{$\sysfnt{1}$}
    \nputbox{-.3,-2.65}{b}{$\sysfnt{2}$}

  \nputbox{1.3,.7}{t}{\color[rgb]{.3,.3,.3}\rule{2pt}{3.8in}}
  \nputbox{-3,.6}{tl}{\large Universal gate set}
  \nputbox{1.4,.7}{tl}{\large \begin{tabular}{@{}l}Postselected\\[-.08in] measurements\end{tabular}}

\end{picture}
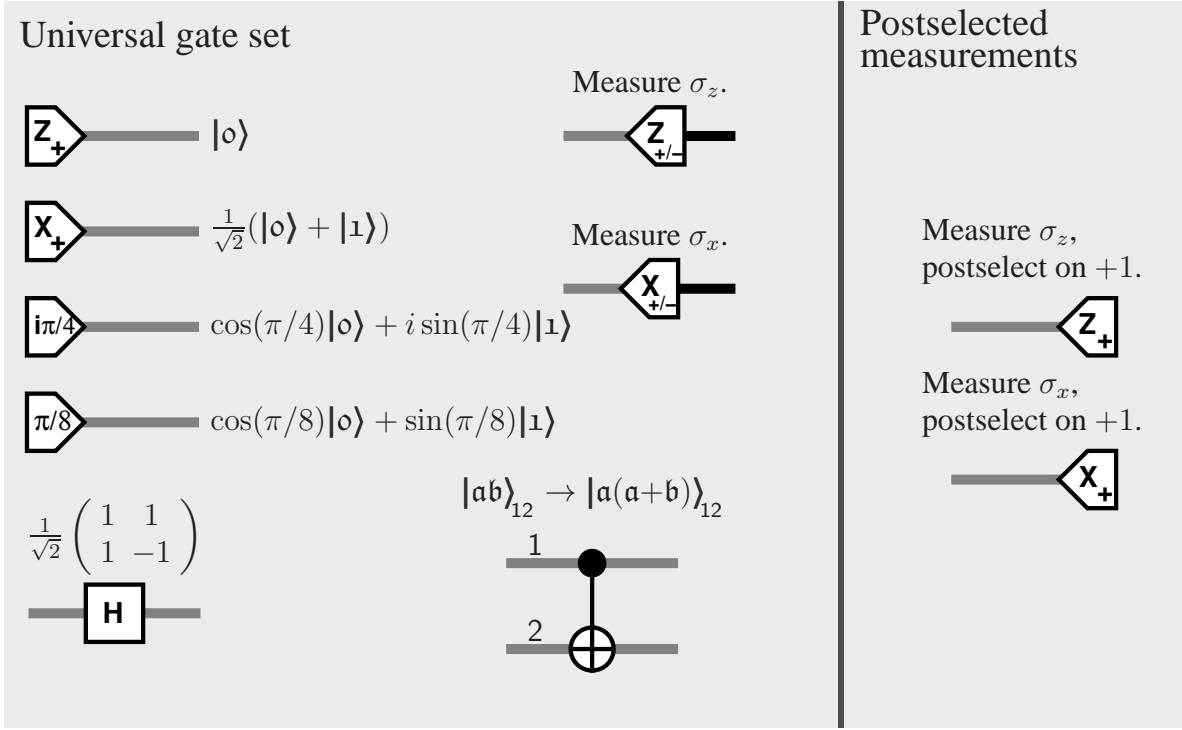
\herefigcap{Network symbols for gates with their definitions.  The
measurement gates' output is classical with value $\biz$ if the
eigenvalue is $+1$ and $\bio$ if it is $-1$. The measurement gates
without output are postselected on the sign of the eigenvalue
indicated in the subscript of the operator symbol ($Z$ or $X$). }
\end{herefig}

An error analysis is not given in this paper. Nevertheless, for making
qualitative observations, it is necessary to specify an error
model. The error model assumes that all errors are probabilistic
Pauli products.  A quantum computation is described in terms of its
quantum network.  Each instance of the quantum computation is modified
by probabilistically inserting Pauli operators after each network
element and before measurements. To describe the constraints on the
probabilities, associate with each network element an \emph{error
location}.  For a state preparation and the Hadamard gate, the error
location is in the qubit line immediately following the state
preparation gate. For the controlled not, it
extends to both qubit lines immediately after the gate. For
measurements, it is on the qubit to be measured just before the
gate. For each Pauli operator $p$ (or product of two Pauli operators
in the case of the cnot) that can act at an error location $l$, there
is an associated error parameter $e_{p,l}$. The strongest assumption
one can make on the error model is that $e_{p,l}$ is the probability
that $p$ acts at location $l$, independently for different
locations. 

One of the distinguishing features of postselected quantum computation
as implemented here is that memory error can be eliminated by
optimistic precomputation. This feature requires that 1. there is no
difficulty in using massively parallel processing to apply gates, and
2. cnots can be applied to any pair of qubits.  But no assumption has
to be made on the relative time taken by cnots and measurements.
Memory error plays a role only when the final state is to be used for
a standard (non-postselected) computation, where it is necessary to
delay until all measurements have been completed before using the
state. For the schemes used here, one measurement delay is sufficient.

\section{An Error-Detecting Code for One Qubit}
\label{sect:an error-detecting}

A four-qubit code can be used for error detection. The simplest such
error-detecting code's stabilizer is generated by $[XXXX],[ZZZZ]$. However,
this encodes two qubits. One of these can be used as the logical
qubit. The other serves as a spectator qubit. It is convenient
not to use a single state for the spectator qubit. Equivalently,
two codes will be used for the logical qubit, depending on the
state of the spectator qubit. The logical and spectator qubits'
operators are chosen to be
\begin{eqnarray*}
 \slb{X}{L} = [XXII] &,& \slb{Z}{L} = [ZIZI]\\
 \slb{X}{S} = [IXIX] &,& \slb{Z}{S} = [IIZZ],
\end{eqnarray*}
where $\sysfnt{L}$ and $\sysfnt{S}$ are labels for the logical and the
spectator qubit, respectively. The convention is that $X$ stands for
$\sigma_x$, $Z$ for $\sigma_z$ and, for example, $[XXII]$ for the
product $\slb{\sigma_x}{1}\slb{\sigma_x}{2}$ when referring to an
operator on four ordered qubits $\sysfnt{1,2,3,4}$.

Every computation begins with preparations of one of two states
defined by $\kets{\biz+}{LS}=\Pi(\slb{Z}{L},\slb{X}{S},[XXXX],[ZZZZ])$
and $\kets{+\biz}{LS}=\Pi(\slb{X}{L},\slb{Z}{S},[XXXX],[ZZZZ])$. The
expression $\Pi(\mathbf{s}_1,\mathbf{s}_2,\ldots)$, where the
$\mathbf{s}_i$ are binary vectors
associated with independent Pauli products $P(\mathbf{s}_i)$,
refers to the state
or subspace of states with eigenvalue $+1\nu(\mathbf{s_i})$ for
$P(\mathbf{s}_i)$. The phase $\nu(\mathbf{s_i})$ is defined
in~\cite{knill:qc2003b} and is always $1$ for the CSS codes used here.
Since both states are pairs of Bell states involving different
pairings of the four qubits, they can be encoded with the networks
shown in Fig.~\ref{fig:codes}

\begin{herefig}
\label{fig:codes}
\begin{picture}(0,2.2)(0,-1.9)
  \nputbox{0,.1}{t}{
     \setlength{\fboxrule}{0pt}\setlength{\fboxsep}{0pt}
     \fcolorbox[rgb]{.92,.92,.92}{.92,.92,.92}{\rule{6in}{0pt}\rule{0pt}{1.9in}}
  }
\nputgr{-.5,0}{tr}{scale=.6}{enc4_zx}\nputbox{-1.7,0}{tr}{\large\textbf{a.}}
\nputgr{+.5,0}{tl}{scale=.6}{enc4_xz}\nputbox{.4,0}{tr}{\large\textbf{b.}}
\end{picture}
\nopagebreak
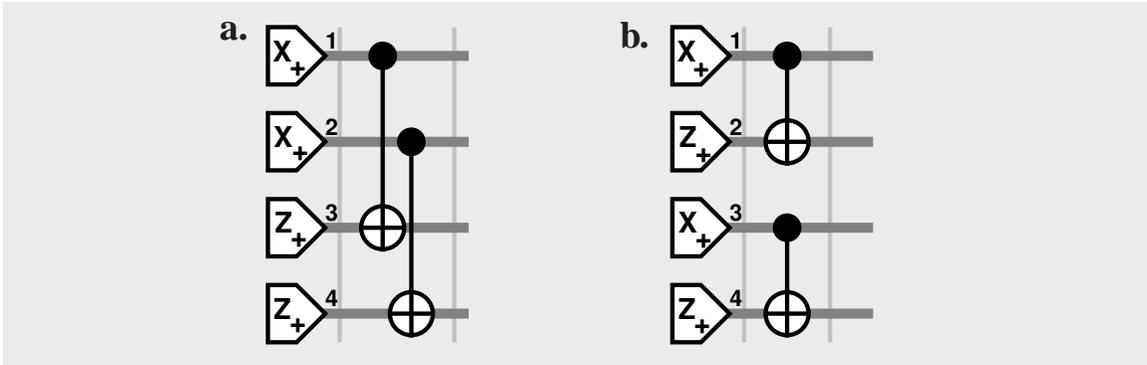
\herefigcap{Fundamental encoding networks.  The stabilizer of the
output of \textbf{a.} is \mbox{generated} by
$[XXXX],[ZZZZ],[ZIZI],[IXIX]$. For the output of \textbf{b.} it is
generated by $[XXXX],[ZZZZ],[XXII],[IIZZ]$. In both cases, the
syndrome is $\mathbf{0}$ (all the eigenvalues are $+1$) with respect to these
generators. It is useful to think of the two states as states of a
logical and a spectator qubit encoded in the stabilizer code
$\Pi([XXXX], [ZZZZ])$ generated by $[XXXX],[ZZZZ]$. This code is
one-error-detecting.  The two encoded qubits are defined by their
encoded Pauli matrices: $\slb{X}{L}=[XXII]$, $\slb{Z}{L}=[ZIZI]$,
$\slb{X}{S}=[IXIX]$, $\slb{Z}{S}=[IIZZ]$. Thus the networks produce
the encoded $\Pi(\slb{Z}{L},\slb{X}{S})$ and
$\Pi(\slb{X}{L},\slb{Z}{S})$ states, respectively. The logical qubit
$\sysfnt{L}$ is used for robustly encoding states after postselection.
Encoded qubit $\sysfnt{S}$ plays a spectator role and is always in an
eigenstate of $\slb{X}{S}$ or $\slb{Z}{S}$. The eigenstate in which it is
intended to be is known and will be used for additional
error detection.}
\end{herefig}

$\kets{\biz+}{LS}$ and $\kets{+\biz}{LS}$ are CSS states, which means
that the stabilizer can be generated by Pauli products that consist
of either all $X$ or all $Z$ operators. This property simplifies
fault-tolerant implementations of encoded operations.  In this case,
encoded cnots and Hadamard gates can be realized transversally and act
on the logical and the spectator qubits in parallel. The code space
$\Pi([XXXX],[ZZZZ])$ is preserved by parallel Hadamard gates of the
four qubits. However, $\slb{X}{L}$ is transformed to an operator
equivalent to $\slb{Z}{S}$ (equivalent with respect to the
stabilizer). This can be fixed by exchanging the middle two
qubits. Thus, to apply simultaneous Hadamard gates to the logical and
the spectator qubits, apply Hadamard gates to each qubit and swap the
middle two qubits. Note that qubit swaps can be done with no error by
doing them only logically. That is, the qubit positions remain fixed,
but their labels are changed. To make this work requires that there is
negligible latency when applying gates between any two qubits, no
matter where they are.

To apply cnots to two encoded logical/spectator qubit pairs, it
suffices to apply four cnots to corresponding pairs of
qubits in the two four-tuples. An example of this occurs in preparing
an encoded Bell state for teleporting gates with error detection. The
network is shown in Fig.~\ref{fig:telestate}.

\begin{herefig}
\label{fig:telestate}
\begin{picture}(0,4)(0,-3.7)
  \nputbox{0,.1}{t}{
     \setlength{\fboxrule}{0pt}\setlength{\fboxsep}{0pt}
     \fcolorbox[rgb]{.92,.92,.92}{.92,.92,.92}{\rule{3.5in}{0pt}\rule{0pt}{3.7in}}
  }
\nputgr{0,0}{t}{scale=.6}{enc4e4}
\end{picture}
\nopagebreak
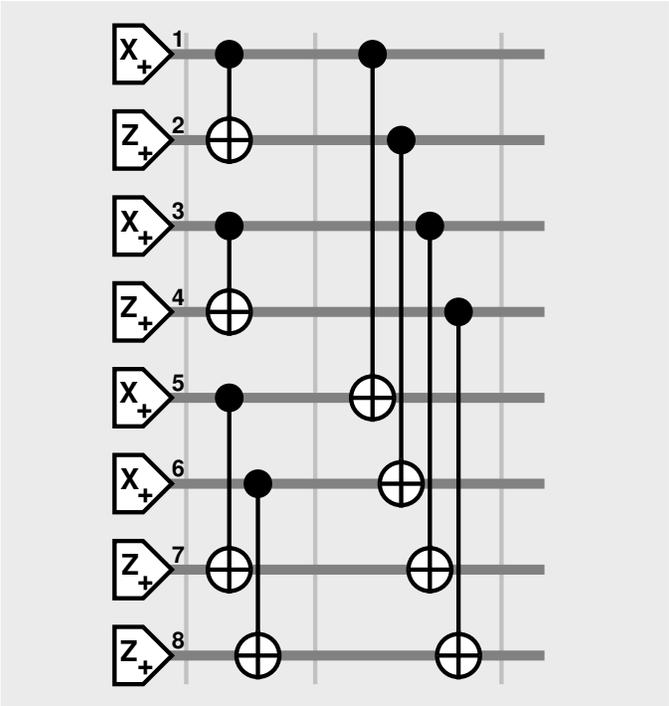
\herefigcap{State used to teleport encoded qubits. Before the final
four cnots, qubits $\sysfnt{1,2,3,4}$ are in the state
$\kets{+\biz}{LS}$.  Qubits
$\sysfnt{5,6,7,8}$ are in the state
$\kets{\biz+}{L'S'}$. The cnots in the last time step
act as encoded cnots on the logical and spectator qubits--a property of
CSS codes. The state after the cnots is
\begin{eqnarray*}
  \begin{array}{l@{}l}
  \Pi(&[XXIIXXII], [ZIZIZIZI], \\
              &[IIZZIIII], [XXXXIIII],[ZZZZIIII] \\
              &[IIIIIXIX], [IIIIXXXX],[IIIIZZZZ]).
  \end{array}
\end{eqnarray*}
This can be recognized as an encoded
$\Pi(\slb{X}{L}\slb{X}{L'},\slb{Z}{L}\slb{Z}{L'},
\slb{Z}{S},\slb{X}{S'})$ state.  In other words, the logical qubits to
be used are entangled in the standard Bell state. The spectator qubits
remain in their original states and are independent of each other.
The light-gray vertical lines separate the different time steps. The
network consists of three steps of parallel operations, including the
state preparation.  The first prepares the qubits, and the second and
third each apply four parallel cnots.  The depth of this network is
three.}
\end{herefig}

\section{Purifying Stabilizer States}
\label{sect:purifying stabilizer}

The state preparations of the previous section can result in errors in
the encoded states at a rate that is proportional to the gate error
rates. For any given four qubits encoding a logical qubit, the
stabilizer is generated either by $[XXXX]$ ,$[ZZZZ]$ and $[IIZZ]$ or
by $[XXXX]$, $ZZZZ]$ and $[IXIX]$, depending on the state of the
spectator qubit. The syndrome for the logical qubit is said to be
$\mathbf{0}$ if the eigenvalues for the appropriate three operators
are all $+1$. The goal is to purify prepared states so as to ensure
that conditionally on the syndromes of the logical qubits being
$\mathbf{0}$, the error in the intended logical state is quadratically
suppressed.

The purification method for Bell states extends to any CSS
state~\cite{duer:qc2002a}. The basic idea is to combine pairs of
states with transversal cnots and measure the qubits of one member
of the pair to determine
whether the syndromes agree.  There are two choices for the
measurement, depending on whether $X$-type or $Z$-type syndromes are to
be compared.  To do both requires two purification stages using
a total of four identically prepared states. Each four-tuple of
corresponding qubits is subjected to one of the two networks
shown in Fig~\ref{fig:zx and xz purify}.

\begin{herefig}
\label{fig:zx and xz purify}
\begin{picture}(0,2.5)(0,-2.2)
  \nputbox{0,.1}{t}{
     \setlength{\fboxrule}{0pt}\setlength{\fboxsep}{0pt}
     \fcolorbox[rgb]{.92,.92,.92}{.92,.92,.92}{\rule{6in}{0pt}\rule{0pt}{2.2in}}
  }
\nputgr{-.5,0}{tr}{scale=.6}{pur1zx}\nputbox{-2.7,0}{tr}{\large\textbf{a.}}
\nputgr{.5,0}{tl}{scale=.6}{pur1xz}\nputbox{.2,0}{tr}{\large\textbf{b.}}
\end{picture}
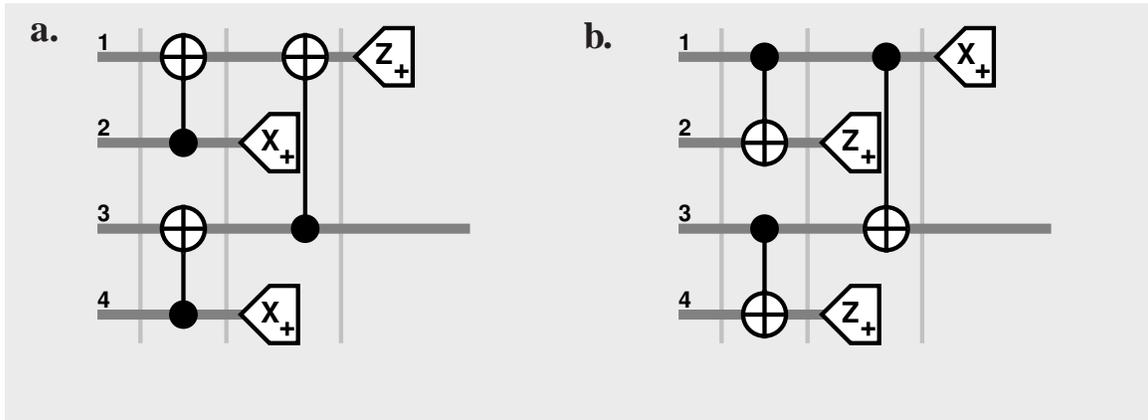
\herefigcap{Purification networks with one-qubit outputs.  The
networks are applied to the corresponding qubits of four identically
prepared CSS states.  The measurement operations are postselected
based on the outcome as shown.  A subscript of $+$ means that only the
$+1$ eigenstate is accepted.  Network $\textbf{a.}$ purifies $X$
syndromes on two pairs of states and then purifies the $Z$ syndromes
in the resulting two states. Network $\textbf{b.}$ purifies $Z$ before
$X$.  }
\end{herefig}

The networks in Fig.~\ref{fig:zx and xz purify} involve postselection
on one measurement outcome for each qubit measurement.  This suffices
for analysis and ensures that if the input states are sufficiently
close to the desired CSS states, the conditionally produced output
state is improved except for local error introduced in the
purification network itself. It is, however, unnecessarily
inefficient. To ensure that if no error occurs, the probability of
success is $1$, the purification can be conditioned on correctness of
appropriately chosen parities of the measurement outcomes.  For
example, the last pairwise combination in Fig.~\ref{fig:zx and xz
purify} checks the $Z$-type syndrome of the two remaining states after
the first round of purification.  If one of the $Z$-type stabilizers
is $\slb{\sigma_z}{1}\slb{\sigma_z}{3}$, then the total parity of the
$Z$ measurements on the first and third set of four corresponding
qubits should be $0$. That is, the outcomes should be both $+1$ or
both $-1$, unless there was an error.

In the concatenated scheme for postselected quantum computation, the
primary states to be prepared are instances of the encoded entangled
state of Fig.~\ref{fig:telestate}. Its preparation with a schematic
purification is shown in Fig~\ref{fig:encoded tele}.

\begin{herefig}
\label{fig:encoded tele}
\begin{picture}(0,4.3)(0,-4.3)
  \nputbox{0,.1}{t}{
     \setlength{\fboxrule}{0pt}\setlength{\fboxsep}{0pt}
     \fcolorbox[rgb]{.92,.92,.92}{.92,.92,.92}{\rule{5in}{0pt}\rule{0pt}{4.3in}}
  }
\nputgr{0,0}{t}{scale=.6}{enctele}
\nputbox{-1.8,-.25}{bl}{\large$B_{zx}$}
\end{picture}
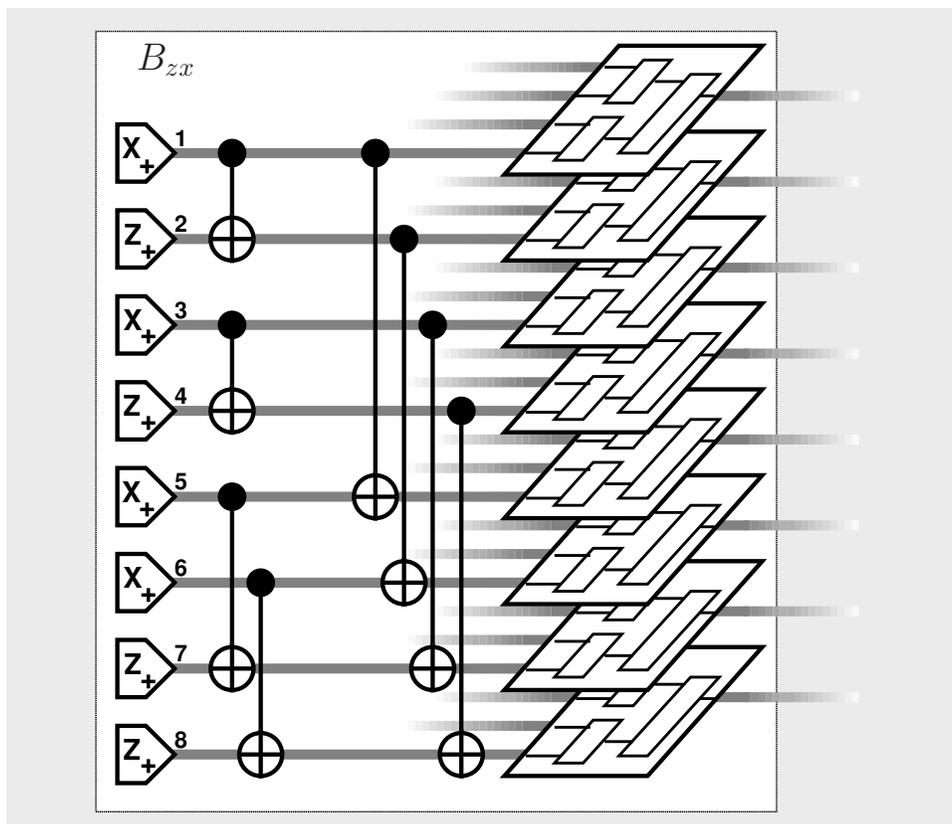
\herefigcap{Typical encoded Bell state preparation with purification.
The spectator qubit is in the state $\Pi(\slb{Z}{S})$
in the top half of the pair of encoded states. In
the bottom half it is in the state $\Pi(\slb{X}{S})$.
The state preparation for one of the four entangled
states that are purified to the one used for teleportation is shown.
Purification occurs in the solid boxes at the right end of the network.
The state preparation network shown is denoted by $B_{zx}$}
\end{herefig}

\section{Encoded Gates with Error Detection}
\label{sect:encoded gates}

As can be seen, the four-qubit error-detecting code requires only
state preparations, measurements in the $Z$ and $X$ basis, and cnots
for the purpose of concatenation. The first task is therefore to show
how to implement these gates in encoded form. One consideration is
that to use prior knowledge of the state of the spectator qubit for
error detection, it is necessary to make sure that it is not affected
by encoded operations and that information about its state can be
extracted from the Bell measurements used for teleportation. To do
this, each teleportation step in an encoded network is labeled
``even'' or ``odd''. The encoded Bell state in even steps (an ``even''
Bell state) has the property that the spectator qubit at the output is
in the state $\Pi(\slb{X}{S})$.  In odd steps, the spectator qubit at
the output is in the state $\Pi(\slb{Z}{S})$.  Thus, the spectator
qubits on the two sides of a transversal Bell measurement used to
teleport are both in the same state and any differences arising from
errors are detected.

Encoded state preparations are shown in Fig.~\ref{fig:encoded state
preps} and involve the use of $B_{zx}$ in either orientation
(producing even or odd encoded Bell states) depending on the
parity of the step.  Encoded measurements are applied directly to the
destination qubits of the state produced by $B_{zx}$ for teleporting
the previous gate. Measuring $X$ on each qubit and postselecting on
$+1$ eigenvalues has the effect of postselecting on the $+1$
$\slb{X}{L}$ eigenstate while rejecting errors affecting the other
$X$-type syndromes. There are one or two such syndromes, depending on
the current state of the spectator qubit, which in turn depends on the
parity of the previous teleportation step.  Measuring $Z$ on each
qubit has an analogous effect.  An encoded cnot is shown in
Fig.~\ref{fig:encoded cnot}.

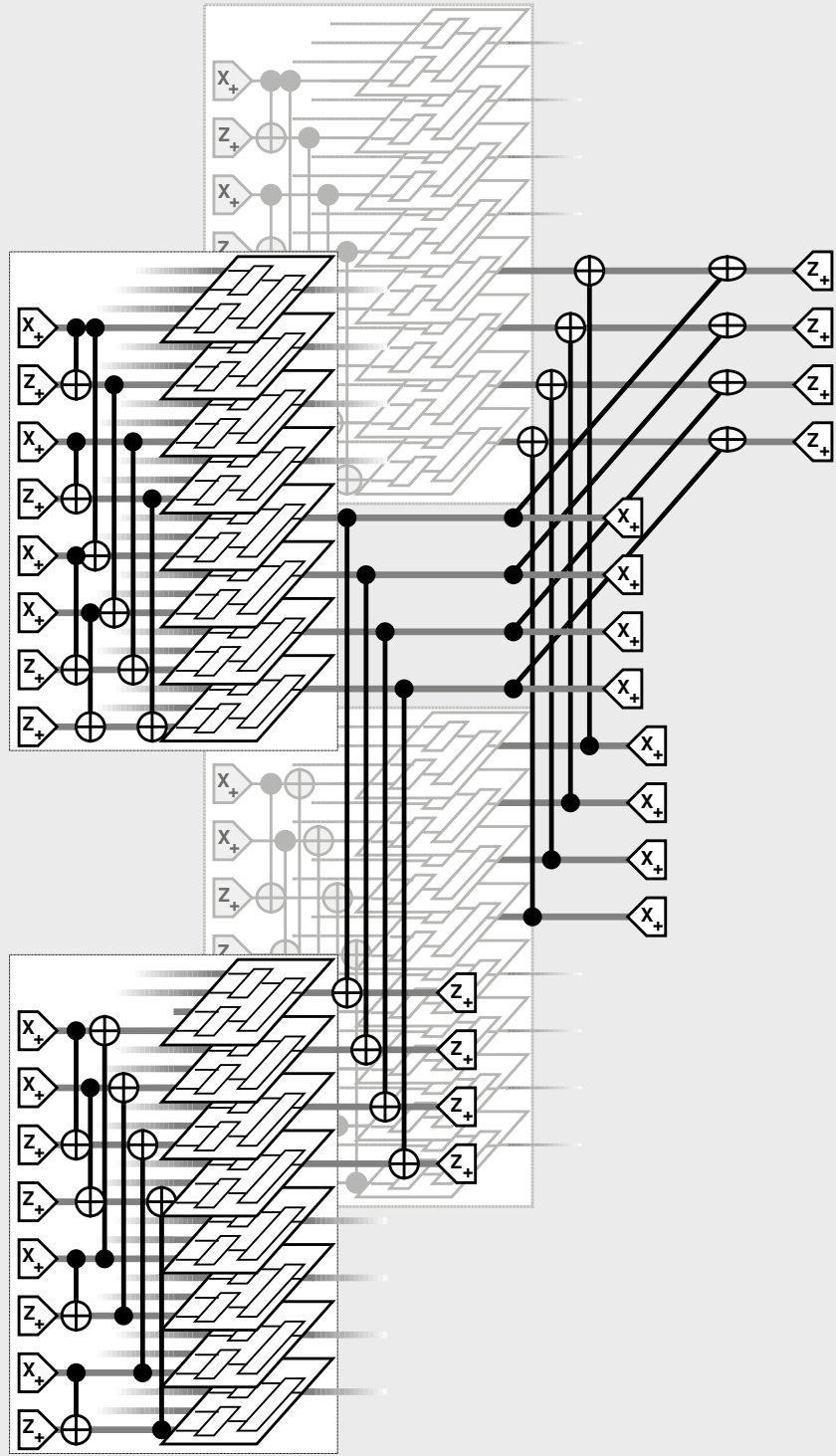
\begin{herefig}
\label{fig:encoded state preps}
\begin{picture}(0,4.2)(0,-4.1)
  \nputbox{0,.1}{t}{
     \setlength{\fboxrule}{0pt}\setlength{\fboxsep}{0pt}
     \fcolorbox[rgb]{.92,.92,.92}{.92,.92,.92}{\rule{6.5in}{0pt}\rule{0pt}{4.1in}}
  }
  \nputgr{-.2,0}{tr}{scale=.6}{encprepx}
  \nputgr{.2,0}{tl}{scale=.6}{encprepz}
  \nputbox{-2.9,0}{tr}{\large\textbf{a.}}
  \nputbox{.4,0}{tr}{\large\textbf{b.}}
\end{picture}
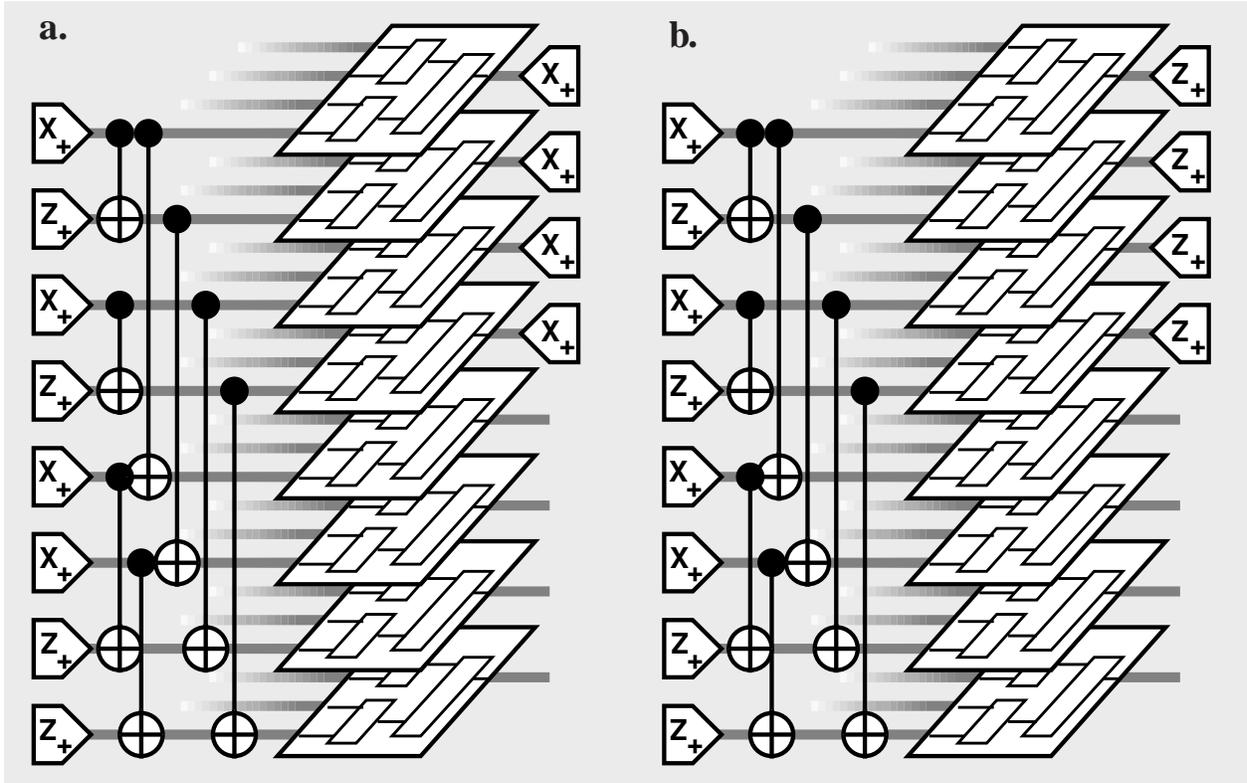
\herefigcap{Encoded preparation of $\sigma_x$ and $\sigma_z$
eigenstates.  Network $\textbf{a.}$ is the $\Pi(\slb{X}{L})$
eigenstate preparation.  Network $\textbf{b.}$ is the
$\Pi(\slb{Z}{L})$ eigenstate preparation.  The networks are shown for
preparations at the zeroth step. The encoded Bell state is therefore
even.  For a preparation requiring an odd Bell state, the measurements
are moved to the bottom four qubits and the top four are the output
qubits.  The postselection ensures that some errors are rejected.  In
$\textbf{a.}$, any errors affecting the $[XXXXIIII]$-syndrome will be
rejected. In $\textbf{b.}$, errors affecting the
$[ZZZZIIII],[IIZZIIII]$-syndrome will be rejected.}
\end{herefig}

\begin{herefig}
\label{fig:encoded cnot}
\begin{picture}(0,7.5)(0,-8)
  \nputbox{0,.1}{t}{
     \setlength{\fboxrule}{0pt}\setlength{\fboxsep}{0pt}
     \fcolorbox[rgb]{.92,.92,.92}{.92,.92,.92}{\rule{6in}{0pt}\rule{0pt}{8in}}
  }
\nputgr{0,0}{t}{scale=.4}{enccnot}
\end{picture}
\herefigcap{Encoded cnot.  The top Bell pairs belong to the
previous operation.  Shown is the cnot with the two sets of
destination qubits (bottom) constituting the second halves of two odd
encoded Bell states.  For the even case, it is necessary
to interchange the preparation
steps for the top and bottom Bell pairs. The cnot is from
the foreground to the background encoded qubits.  The idea is to apply
a cnot transversally, then perform the error-detecting Bell
measurements. The Bell measurements consist of cnots followed
by $X$ and $Z$ measurements on the control and target qubit, respectively.
To maximize the ability to optimistically precompute and
parallelize, the cnots needed for the Bell measurements have
been arranged so that they can be interchanged with the transversal
cnots that realize the operation. This delays the eight
measurements on the input (top) qubits by one step, but allows the
other eight measurements to be performed immediately.
}
\end{herefig}

The postselection shown in Fig.~\ref{fig:encoded cnot} ensures that
single errors and some double errors are eliminated. It also ensures
that the Pauli corrections normally needed in teleportation are not
required, since the Bell measurements project onto the standard Bell
state ${1\over\sqrt{2}}(\ket{\biz\biz}+\ket{\bio\bio})$ given
postselection. Efficiency can be gained by postselecting on parities
of the measurement associated with the syndromes and keeping track of
the Pauli corrections required. The actual implementation of the
correction can typically be deferred to the very end of a large state
preparation procedure or can just be kept track of by bookkeeping
techniques (see \cite{steane:qc2002a}).

\section{Encoding Arbitrary One-Qubit States}
\label{sect:encoding arbitrary}

In order to enable non-Clifford gates at the top encoded level
of the concatenated error-detecting code, it is necessary
to start with (noisy) non-stabilizer one-qubit states and encode
them into the top level. This must be done in such a way
that excess error in the encoded state compared to the starting
state is well bounded. This is accomplished by teleporting
it into the code by use of a prepared entanglement between a qubit
and a logical qubit as shown in Fig.~\ref{fig:encoding arbitrary}.
An alternative method for preparing the required entanglement,
by decoding one half of a fully encoded Bell pair, is shown
in Fig.~\ref{fig:encoding arbitrary (dec)}.

\begin{herefig}
\label{fig:encoding arbitrary}
\begin{picture}(0,4)(-1,-4)
  \nputbox{-1.2,.1}{t}{
     \setlength{\fboxrule}{0pt}\setlength{\fboxsep}{0pt}
     \fcolorbox[rgb]{.92,.92,.92}{.92,.92,.92}{\rule{6in}{0pt}\rule{0pt}{4in}}
  }
  \nputgr{0,0}{t}{scale=.6}{psienc}
  \nputgr{-2.5,-.75}{tr}{scale=.6}{psienc_v1}
   \nputbox{.61,-.1}{t}{\small $\ket{\psi}$}
\end{picture}
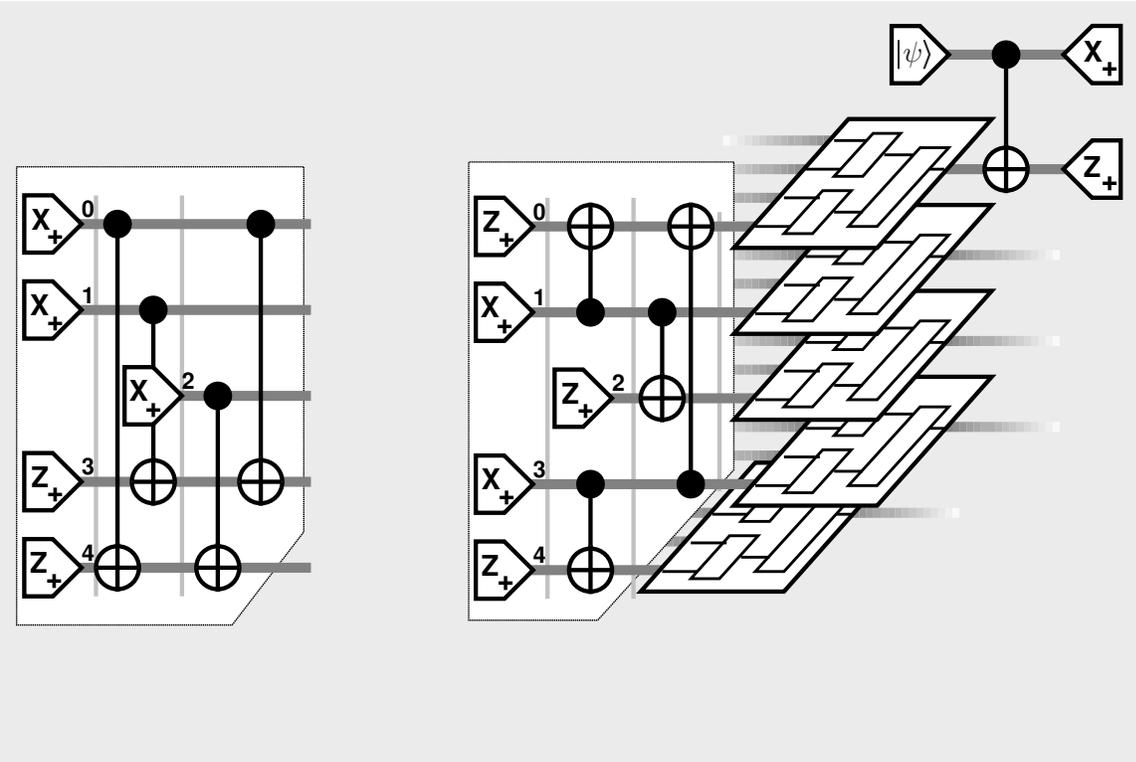
\herefigcap{Encoding an arbitrary state to the next level.  The full
network on the right first entangles qubit $\sysfnt{0}$ with
the logical qubit encoded in qubits
$\sysfnt{1},\sysfnt{2},\sysfnt{3},\sysfnt{4}$.  The spectator qubit is
in the state $\Pi(\slb{Z}{S})$, suitable for an even output in a large
network.  The prepared entanglement is purified. Since the resulting
state is defined by $\Pi([IXXXX],[IZZZZ],[IIIZZ],[XXXII],[ZZIZI])$, it
is one-error detecting.  There is an unavoidable memory delay on qubit
$\sysfnt{4}$.  It may be chosen to occur either before or after
purification.  The last step is to teleport the desired state
$\ket{\psi}$ into the code.  The state preparation before purification
for having the spectator qubit in state $\Pi(\slb{X}{S})$ is shown on
the left without the subsequent purification and teleportation
networks. This can be used for odd steps. In this case, the
unavoidable memory delay is on qubit $\sysfnt{1}$.  }
\end{herefig}

\begin{herefig}
\label{fig:encoding arbitrary (dec)}
\begin{picture}(0,5)(0,-5)
  \nputbox{0,.1}{t}{
     \setlength{\fboxrule}{0pt}\setlength{\fboxsep}{0pt}
     \fcolorbox[rgb]{.92,.92,.92}{.92,.92,.92}{\rule{6in}{0pt}\rule{0pt}{5in}}
  }
  \nputgr{0,0}{t}{scale=.6}{psienc_v2}
  \nputbox{1.44,-.1}{t}{\small $\ket{\psi}$}
  \nputbox{.45,-.52}{tl}{$D_{z}$}
  \nputgr{1.25,-2.8}{tl}{scale=.6}{psienc_v3}
  \nputbox{1.7,-2.87}{tl}{$D_{x}$}
\end{picture}
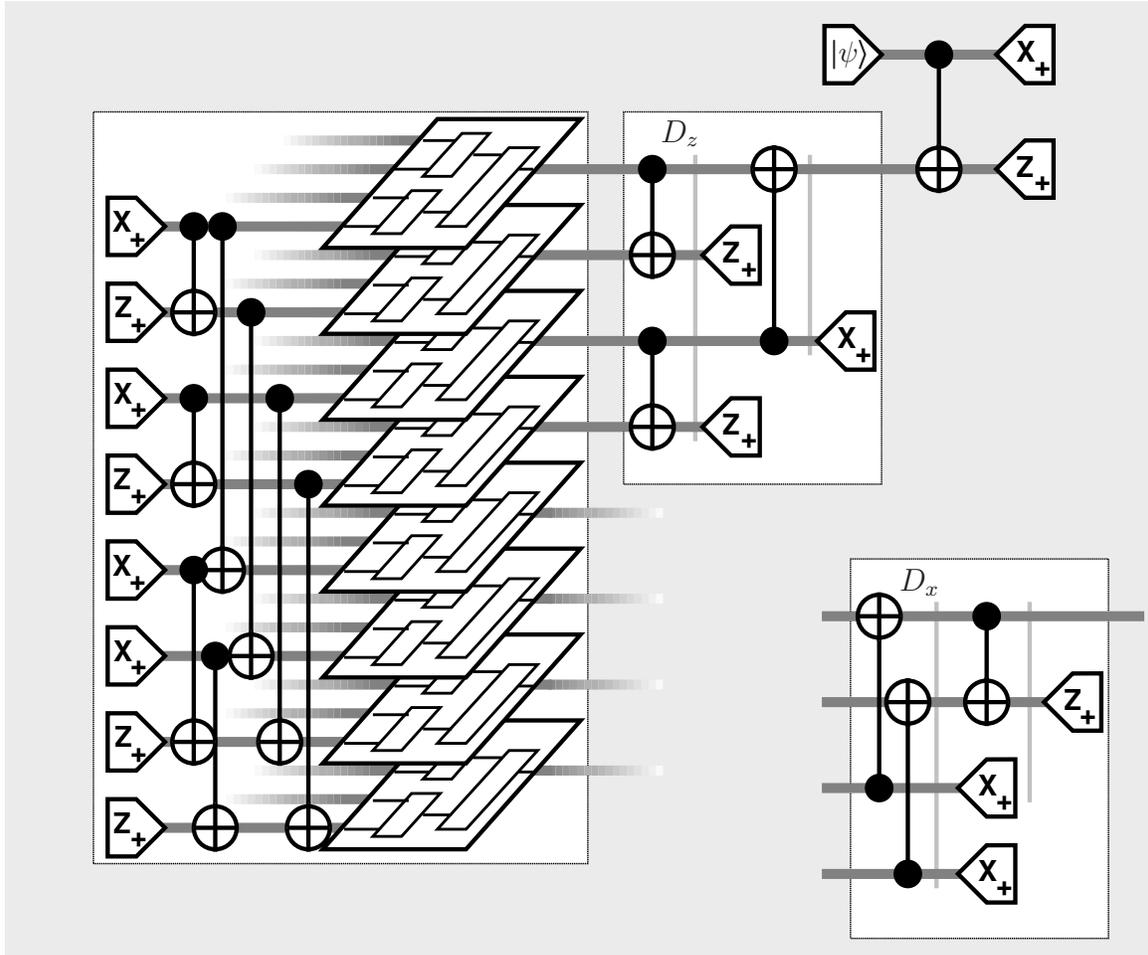
\herefigcap{Encoding an arbitrary state using an encoded Bell state.
Here, one of the logical qubits in an encoded Bell state is first
decoded, then used to teleport $\ket{\psi}$ into the code.  The full
procedure is shown for decoding a logical qubit where its spectator
qubit is in the $\Pi(\slb{Z}{S})$ state. The decoding network is
labeled $D_z$. Note that it is identical to a version of the
purification network~\ref{fig:zx and xz purify}. To decode with the
spectator qubit in the $\Pi(\slb{X}{S})$ state (the bottom half of the
encoded Bell pair in the picture), use the network labeled $D_x$. The
decoding network is designed so that the three measured qubits yield
syndrome information. Any error that affects the syndrome is detected
and results in rejection of the state obtained. Note that memory,
given optimistic precomputation on other qubits, delays do not occur.
}
\end{herefig}

\section{Concatenation}
\label{sect:concatenation}

The techniques of the previous sections can be concatenated by the
usual method of substituting qubits by logical qubits and gates by
encoded gates based on the networks shown.  Because of the pervasive
use of teleported operations and postselection, it can be seen that
the maximum depth of the networks obtained is determined by the depth
of the network at the first level of concatenation. Counting state
preparation and measurement, this depth is at most $8$ for CSS gates
and $9$ for encoding an arbitrary state, which can be compared to the
circuits of depth $5$ for postselected quantum computation found
in~\cite{fenner:qc2003a}.

The concatenated circuits can be simplified by avoiding explicit
concatenation of gates involved in the Bell state measurements
and the encoded cnots. That is, instead of reimplementing
the cnots required for these two operations, they can be directly applied
transversally to the concatenated error-detecting code, as can
the subsequent measurements.

\section{One-Qubit States for Universal Computation}
\label{sect:purifying one-qubit}

In order to enable universal quantum computation at the top encoded
level, one can use a scheme for encoding a few additional one-qubit
states with bounded error.  They can then be purified by using the
ability to implement nearly error-free CSS gates.  For real quantum
computation, it suffices to prepare the state
$\ket{\pi/8}=\cos(\pi/8)\ket{\biz}+\sin(\pi/8)\ket{\bio}$, the $+1$
eigenstate of the Hadamard gate.  However, it is convenient to have
the ability to use complex phases, and for that it suffices to use the
state $\ket{i\pi/4}={1\over\sqrt{2}}(\ket{\biz}+i\ket{\bio})$, the
$+i$ eigenstate of $\sigma_x$ followed by $\sigma_z$. That these suffice is
due to the networks shown in Fig~\ref{fig:state to rots}.  They make
it possible to implement the rotations $e^{-i\sigma_x\pi/4}$ and
$e^{-i\sigma_y\pi/8}$. Together with the CSS gates they form a
universal set~\cite{knill:qc1997a}.

\begin{herefig}
\label{fig:more gates}
\begin{picture}(0,1.9)(0,-1.8)
  \nputbox{0,.1}{t}{
     \setlength{\fboxrule}{0pt}\setlength{\fboxsep}{0pt}
     \fcolorbox[rgb]{.92,.92,.92}{.92,.92,.92}{\rule{3.3in}{0pt}\rule{0pt}{1.8in}}
  }
  \nputgr{0,0}{t}{scale=.6}{yop}
  \nputbox{-.3,-.35}{c}{\large $=$}
  \nputbox{-.3,-1.2}{c}{\large $=$}
\end{picture}
\herefigcap{
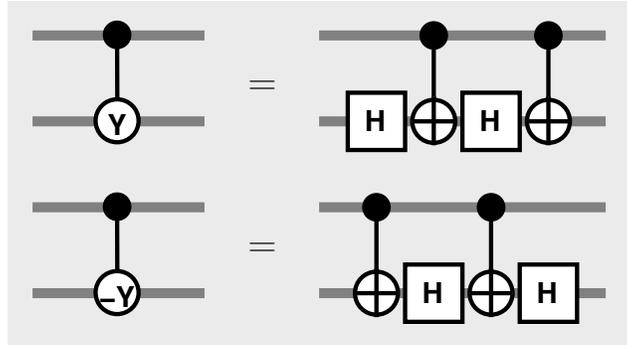
Definition of controlled-$Y$ gates.
}
\end{herefig}

\begin{herefig}
\label{fig:state to rots}
\begin{picture}(0,2.1)(0,-2)
  \nputbox{0,.1}{t}{
     \setlength{\fboxrule}{0pt}\setlength{\fboxsep}{0pt}
     \fcolorbox[rgb]{.92,.92,.92}{.92,.92,.92}{\rule{6.6in}{0pt}\rule{0pt}{2in}}
  }
  \nputgr{-.3,0}{tr}{scale=.6}{rxpi2}
  \nputbox{-1.75,-.65}{c}{\large $=$}
  \nputbox{-1.75,-1.7}{c}{\large $=$}
  \nputgr{.35,0}{tl}{scale=.6}{rypi4}
  \nputbox{1.8,-.65}{c}{\large $=$}
  \nputbox{1.8,-1.7}{c}{\large $=$}
\end{picture}
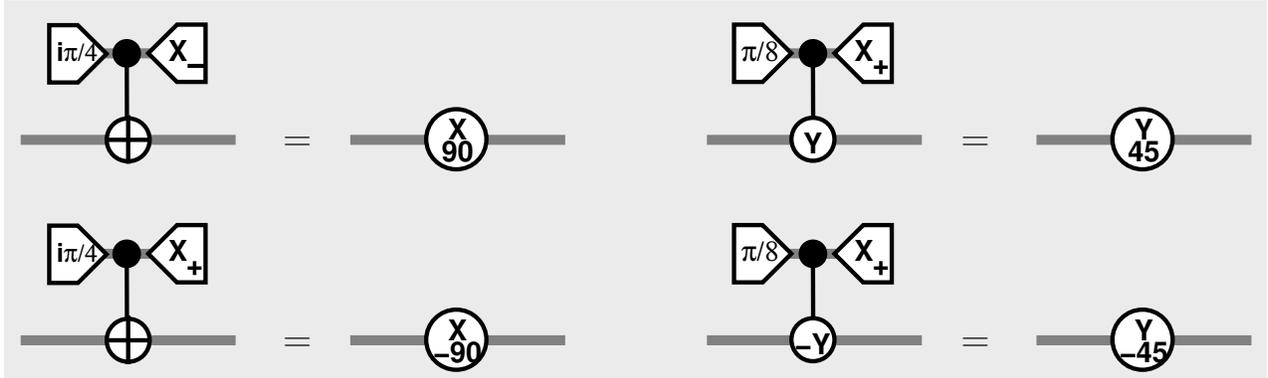
\herefigcap{Implementation of $e^{-i\sigma_x\pi/4}$ and
$e^{-i\sigma_y\pi/8}$ rotations. Note that postselection based on
measurement outcomes is used.  The sign of the rotation depends on the
measurement outcome, so it is possible, in principle, to implement the
rotations without postselection by following with $\pm 180\dg$ or $\pm
90\dg$ rotations if the opposite measurement outcome is obtained.  }
\end{herefig}

There are several ways in which $\ket{i\pi/4}$ and $\ket{\pi/8}$ can
be purified by use of CSS gates. Consider $\ket{i\pi/4}$. This is the $-i$
eigenstate of $\sigma_x\sigma_z$.  Thus a measurement to verify
$\ket{i\pi/4}$ can be performed using two instances of $\ket{i\pi/4}$
and a controlled-$\sigma_x\sigma_z$ gate, as shown in
Fig.~\ref{fig:purify states}.

\begin{herefig}
\label{fig:purify states}
\begin{picture}(0,1.3)(0,-1.2)
  \nputbox{0,.1}{t}{
     \setlength{\fboxrule}{0pt}\setlength{\fboxsep}{0pt}
     \fcolorbox[rgb]{.92,.92,.92}{.92,.92,.92}{\rule{4in}{0pt}\rule{0pt}{1.2in}}
  }
\nputgr{0,-.1}{t}{scale=.6}{pi4purify}
\end{picture}
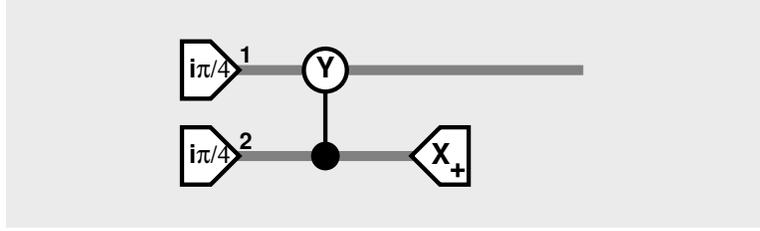
\herefigcap{Purifying $\ket{i\pi/4}$ using
cnots and Hadamard gates. The controlled gate
shown is defined in Fig.~\ref{fig:more gates} and involves two cnots and
Hadamard gates. Any single error is detected by the $X$
measurement and rejected by postselection.}
\end{herefig}

The conditional gate in Fig.~\ref{fig:purify states} kicks back a
phase of $-i$ on $\kets{\bio}{2}$, which changes $\kets{i\pi/4}{2}$ to
$\kets{+}{2}={1\over\sqrt{2}}(\kets{\biz}{2}+\kets{\bio}{2})$. The
measurement therefore succeeds.  By decohering the input states using
random applications of $i\sigma_y$ (if necessary), the error on the
two prepared states can be assumed to be 
random $\sigma_x$ noise.  It can be seen that a $\slb{\sigma_x}{1}$
after the state preparation is equivalent to
$\slb{\sigma_x}{1}\slb{\sigma_z}{2}$ before the measurement. The
measurement outcome therefore changes.  A $\slb{\sigma_x}{2}$ error is
equivalent to $\slb{\sigma_y}{1}\slb{\sigma_y}{2}$ before the
measurement and again, the measurement outcome changes. This shows
that any single error is rejected by the network.  Therefore, the error in the
prepared $\ket{i\pi/4}$ is reduced quadratically given success.

To purify $\ket{\pi/8}$ using CSS gates is more difficult.
First note that a measurement to verify $\ket{\pi/8}$ can
be implemented with a conditional-Hadamard gate, which in
turn requires two prepared $\ket{\pi/8}$ states as
shown in Fig.~\ref{fig:chad}

\begin{herefig}
\label{fig:chad}
\begin{picture}(0,1.6)(0,-1.5)
  \nputbox{0,.1}{t}{
     \setlength{\fboxrule}{0pt}\setlength{\fboxsep}{0pt}
     \fcolorbox[rgb]{.92,.92,.92}{.92,.92,.92}{\rule{6.5in}{0pt}\rule{0pt}{1.5in}}
  }
  \nputgr{0,-.1}{t}{scale=.6}{chad}
  \nputbox{-1.3,-.75}{c}{\large $=$}
  \nputbox{1,-.75}{c}{\large $=$}
  \nputbox{3.1,-.6}{c}{$\ket{\pi/8}$}
\end{picture}
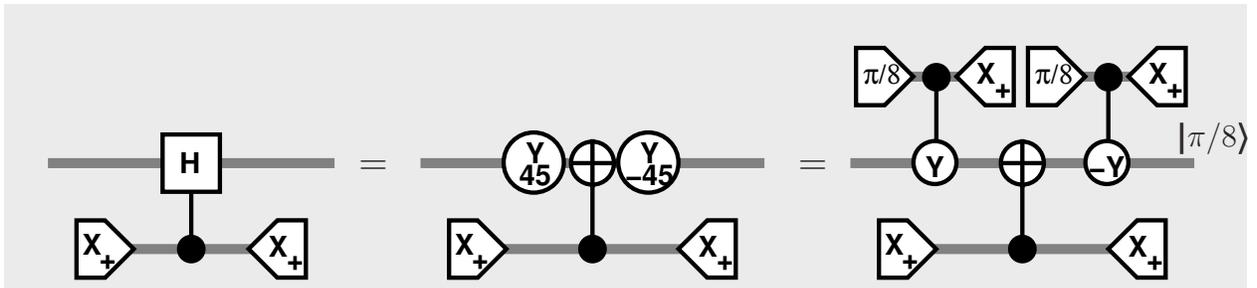
\herefigcap{Measurement to project the input onto $\ket{\pi/8}$. If
the $X$ measurement results in the $-1$ eigenstate, then the
measurement projects the input onto the orthogonal state.}
\end{herefig}

By randomly applying Hadamard gates to the prepared $\ket{\pi/8}$
states, the error model can be assumed to be probabilistic $\sigma_y$
errors. That is, after an error, the state prepared is the orthogonal
state $\ket{5\pi/8}$. For the $e^{-i\sigma_y\pi/8}$ rotations
implemented with prepared $\ket{\pi/8}$'s, the error causes an
unintentional additional $\sigma_y$. In the context of the measurement
circuit of Fig.~\ref{fig:chad}, such an error in the second $Y$
rotation results in a $\sigma_y$ error at the output.  In the first
$Y$ rotation, its effect is to introduce a $\sigma_y$ error at the
input. 

A straightforward way to purify the $\ket{\pi/8}$ state is to use a
self-dual CSS code for encoding one qubit with good error-detecting
properties. An example is the 7-qubit Hamming code.  Specifically,
consider a CSS code for one qubit for which the stabilizer is
invariant when Hadamard gates are applied to each qubit (transversal
Hadamard) and such that the logical $X$ and $Z$ operators for the
encoded qubit are exchanged under the transversal Hadamard. Such codes can
be obtained from self-dual classical codes
(see~\cite{steane:qc2003a}).  Then one can encode the logical
$\kets{\biz}{L}$ with CSS gates. A measurement of the encoded Hadamard
operator is performed by using an ancilla qubit prepared in $\ket{+}$
and conditional Hadamard gates from the ancilla to each qubit of the
code. An $X$ measurement with postselection on $+1$ should project onto
the encoded $\ket{\pi/8}$ state unless errors occur in the
$\ket{\pi/8}$ state preparations. Errors in the conditional Hadamard
gates cause syndrome changes via the $\sigma_y$ effects on the qubits
of the code.  These syndrome changes are detected when the state is
decoded to extract the encoded $\ket{\pi/8}$. Conditionally on no
detected errors, the $\ket{\pi/8}$ has much reduced error. Self-dual
CSS codes with the ability to detect $\sigma_y$ errors at a rate above
$10\;\%$ exist. (This is because classical self-dual codes that meet the
Gilbert-Varshamov bound exist, see~\cite{macwilliams:qc1977a}, chapter
19, and~\cite{calderbank:qc1996b}).

\section{Preparing Arbitrary Stabilizer States}
\label{sect:preparing arbitrary}

Given a stabilizer state defined by generator matrix $Q$, there is an
encoding network using CSS gates and $e^{-i\sigma_x\pi/4}$
rotations~\cite{cleve:qc1996b}. Using the above methods, it is
possible to use postselected computation to prepare the state $\Pi(Q)$
at a level of concatenation where the postselected, encoded CSS gates
are sufficiently accurate to also enable accurate $90\dg$ $\sigma_x$
rotations with $\ket{i\pi/4}$ preparations.  Once the encoded $\Pi(Q)$
state has been obtained, one can apply the decoding networks $D_x$ and
$D_z$ shown in Fig.~\ref{fig:encoding arbitrary (dec)} to collapse the
concatenated code to one qubit. The collapse should be performed
bottom up, not top down. That is, decoding is performed on physical
qubits four at a time until only one qubit remains for each top-level
logical qubit. Conditionally on no errors being detected during the
decoding, one would expect that the resulting state is perturbed from
$\Pi(Q)$ only by local errors that either had not been detected
when the encoded state was prepared, or were introduced during decoding
without being detected. Provided the local error-rate is small enough,
the state can be used in a general scheme for fault-tolerant quantum
computation~\cite{knill:qc2003b}.

It is worth noting again that with optimistic precomputation, memory
errors do not play a role until the state is used elsewhere in a
non-postselected computation. That is, every qubit is subject to an
operation at every step.  Before the state can be used in a
non-postselected computation, there is one memory delay, which is required to
wait for a positive outcome of the last measurement in the decoding
circuit.  The price of avoiding memory delays is massive parallelism.

\section{Discussion}
\label{sect:discussion}

At this point, the only evidence that the methods above lead to
fault-tolerant postselected quantum computation with reasonable error
thresholds is qualitative. Inspection of the methods suggests that
isolated errors within each code block will be detected so that,
conditionally on success within one level of concatenation, the
encoded error rate is expected to be proportional to the square of the
base error rate.  Compared to the concatenated codes typically used
for regular quantum computation, the codes here are significantly
simpler.  The small depth and breadth of the relevant circuits
suggests that the constant of proportionality should be relatively
small.  However, it is necessary to take into account that
postselection requires conditioning on potentially rare
events. Determining whether the relative probabilities of error
combinations really behave well in this setting requires further
analysis.

As shown, accurate postselected quantum computation is inefficient.
The following can be used to improve the overall efficiency.  First,
it is a good idea to stop concatenation using short codes as soon as
possible. In an application to state preparation, the main goal is to
achieve sufficient accuracy for other, non-postselected techniques
that work well at lower error rates can be used to scale up. This
approach is used in~\cite{knill:qc2003b} in the context of erasure
errors to explicitly show efficiency of a fault-tolerant
scheme. Second, wherever possible, the postselection should be
modified to have probability of success $1$ if there are no errors. In
the context of the methods used here, this requires using bookkeeping
techniques to to keep track of the syndromes and adjust the
postselection accordingly.  Finally, the small depth of the networks
makes it possible to combine postselected subnetworks in a tree-like
fashion. This avoids having to redo the entire preparation each time
and, under ideal circumstances, makes the total overhead for a
successful computation efficient.  How efficiently one can prepare
states with the schemes given in this paper using the improvements
just mentioned remains to be determined.

The code and networks used here have the remarkable feature of
involving little more than Bell state preparation followed by a small
number of different ways of combining them and partial and complete
Bell state measurements. It might be useful to better understand the
role of Bell states and Bell measurements as used here.  In this
context, one could ask whether a smaller error-detecting code could be
used. Unfortunately, there are no three-qubit error-detecting codes
for qubits~\cite{grassl:1996a}. However, there are such codes for
qutrits (three-level quantum systems)~\cite{cleve:qc1999a}, suggesting that the
error-tolerance for qutrits might be more favorable.  This is worth
investigating, but it is necessary to be cautious when comparing error
rates for qubit and qutrit Clifford gates.  At least some of the
latter are physically more complex, meaning that a higher error rate
is to be expected. For example, the analogue of the Hadamard gate
for qutrits is not readily implemented in spin-$1$ systems by use of
typically available control based on spin observables. Multiple
individually addressed level-transitions are needed to implement it.
They also involve a large-dimensional space, with more opportunities
for errors to occur. Furthermore, the four-qubit error-detecting
code is in a sense more compact then the three-qutrit error-detecting
code. The former lives in a 16 dimensional state space, the latter in
a 27 dimensional one.

\begin{acknowledgments}
This work was supported by the U.S. National Security Agency. It is a
contribution of the National Institute of Standards and Technology, an
agency of the U.S. government, and is not subject to U.S. copyright.
\end{acknowledgments}

\bibliography{journalDefs,qc,state}

\begin{thebibliography}{28}
\expandafter\ifx\csname natexlab\endcsname\relax\def\natexlab#1{#1}\fi
\expandafter\ifx\csname bibnamefont\endcsname\relax
  \def\bibnamefont#1{#1}\fi
\expandafter\ifx\csname bibfnamefont\endcsname\relax
  \def\bibfnamefont#1{#1}\fi
\expandafter\ifx\csname citenamefont\endcsname\relax
  \def\citenamefont#1{#1}\fi
\expandafter\ifx\csname url\endcsname\relax
  \def\url#1{\texttt{#1}}\fi
\expandafter\ifx\csname urlprefix\endcsname\relax\def\urlprefix{URL }\fi
\providecommand{\bibinfo}[2]{#2}
\providecommand{\eprint}[2][]{\url{#2}}

\bibitem[{\citenamefont{Knill}(2003)}]{knill:qc2003b}
\bibinfo{author}{\bibfnamefont{E.}~\bibnamefont{Knill}} (\bibinfo{year}{2003}),
  \bibinfo{note}{quant-ph/0312190}.

\bibitem[{\citenamefont{D\"ur and Briegel}(2003)}]{duer:qc2002a}
\bibinfo{author}{\bibfnamefont{W.}~\bibnamefont{D\"ur}} \bibnamefont{and}
  \bibinfo{author}{\bibfnamefont{H.-J.} \bibnamefont{Briegel}},
  \bibinfo{journal}{Phys. Rev. Lett.} \textbf{\bibinfo{volume}{90}},
  \bibinfo{pages}{067901/1} (\bibinfo{year}{2003}).

\bibitem[{\citenamefont{Steane and Ibinson}(2003)}]{steane:qc2003a}
\bibinfo{author}{\bibfnamefont{A.~M.} \bibnamefont{Steane}} \bibnamefont{and}
  \bibinfo{author}{\bibfnamefont{B.}~\bibnamefont{Ibinson}}
  (\bibinfo{year}{2003}), \bibinfo{note}{quant-ph/0311014}.

\bibitem[{\citenamefont{Shor}(1996)}]{shor:qc1996a}
\bibinfo{author}{\bibfnamefont{P.~W.} \bibnamefont{Shor}}, in
  \emph{\bibinfo{booktitle}{Proceedings of the 37th Symposium on the
  Foundations of Computer Science (FOCS)}} (\bibinfo{publisher}{IEEE press},
  \bibinfo{address}{Los Alamitos, California}, \bibinfo{year}{1996}), pp.
  \bibinfo{pages}{56--65}.

\bibitem[{\citenamefont{Kitaev}(1997)}]{kitaev:qc1996a}
\bibinfo{author}{\bibfnamefont{A.~Y.} \bibnamefont{Kitaev}}, in
  \emph{\bibinfo{booktitle}{Quantum Communication and Computing and
  Measurement}}, edited by \bibinfo{editor}{\bibfnamefont{O.~H.}
  \bibnamefont{et~al.}} (\bibinfo{publisher}{Plenum}, \bibinfo{address}{New
  York}, \bibinfo{year}{1997}).

\bibitem[{\citenamefont{Knill and Laflamme}(1996)}]{knill:qc1996b}
\bibinfo{author}{\bibfnamefont{E.}~\bibnamefont{Knill}} \bibnamefont{and}
  \bibinfo{author}{\bibfnamefont{R.}~\bibnamefont{Laflamme}},
  \bibinfo{type}{Tech. Rep.} \bibinfo{number}{LAUR-96-2808},
  \bibinfo{institution}{Los Alamos National Laboratory} (\bibinfo{year}{1996}),
  \bibinfo{note}{quant-ph/9608012}.

\bibitem[{\citenamefont{Aharonov and Ben-Or}(1996)}]{aharonov:qc1996a}
\bibinfo{author}{\bibfnamefont{D.}~\bibnamefont{Aharonov}} \bibnamefont{and}
  \bibinfo{author}{\bibfnamefont{M.}~\bibnamefont{Ben-Or}}, in
  \emph{\bibinfo{booktitle}{Proceedings of the 29th Annual ACM Symposium on the
  Theory of Computation (STOC)}} (\bibinfo{publisher}{ACM Press},
  \bibinfo{address}{New York, New York}, \bibinfo{year}{1996}), pp.
  \bibinfo{pages}{176--188}.

\bibitem[{\citenamefont{Aharonov and Ben-Or}(1999)}]{aharonov:qc1999a}
\bibinfo{author}{\bibfnamefont{D.}~\bibnamefont{Aharonov}} \bibnamefont{and}
  \bibinfo{author}{\bibfnamefont{M.}~\bibnamefont{Ben-Or}}
  (\bibinfo{year}{1999}), \bibinfo{note}{quant-ph/9906129}.

\bibitem[{\citenamefont{Knill et~al.}(1998{\natexlab{a}})\citenamefont{Knill,
  Laflamme, and Zurek}}]{knill:qc1997a}
\bibinfo{author}{\bibfnamefont{E.}~\bibnamefont{Knill}},
  \bibinfo{author}{\bibfnamefont{R.}~\bibnamefont{Laflamme}}, \bibnamefont{and}
  \bibinfo{author}{\bibfnamefont{W.}~\bibnamefont{Zurek}},
  \bibinfo{journal}{Proc. R. Soc. Lond. A} \textbf{\bibinfo{volume}{454}},
  \bibinfo{pages}{365} (\bibinfo{year}{1998}{\natexlab{a}}).

\bibitem[{\citenamefont{Gottesman}(1998)}]{gottesman:qc1997a}
\bibinfo{author}{\bibfnamefont{D.}~\bibnamefont{Gottesman}},
  \bibinfo{journal}{Phys. Rev. A} \textbf{\bibinfo{volume}{57}},
  \bibinfo{pages}{127} (\bibinfo{year}{1998}).

\bibitem[{\citenamefont{Knill et~al.}(1998{\natexlab{b}})\citenamefont{Knill,
  Laflamme, and Zurek}}]{knill:qc1998a}
\bibinfo{author}{\bibfnamefont{E.}~\bibnamefont{Knill}},
  \bibinfo{author}{\bibfnamefont{R.}~\bibnamefont{Laflamme}}, \bibnamefont{and}
  \bibinfo{author}{\bibfnamefont{W.~H.} \bibnamefont{Zurek}},
  \bibinfo{journal}{Science} \textbf{\bibinfo{volume}{279}},
  \bibinfo{pages}{342} (\bibinfo{year}{1998}{\natexlab{b}}).

\bibitem[{\citenamefont{Zalka}(1996)}]{zalka:qc1996a}
\bibinfo{author}{\bibfnamefont{C.}~\bibnamefont{Zalka}} (\bibinfo{year}{1996}),
  \bibinfo{note}{quant-ph/9612028}.

\bibitem[{\citenamefont{Preskill}(1998)}]{preskill:qc1998a}
\bibinfo{author}{\bibfnamefont{J.}~\bibnamefont{Preskill}},
  \bibinfo{journal}{Proc. R. Soc. Lond. A} \textbf{\bibinfo{volume}{454}},
  \bibinfo{pages}{385} (\bibinfo{year}{1998}).

\bibitem[{\citenamefont{Steane}(1999)}]{steane:qc1999a}
\bibinfo{author}{\bibfnamefont{A.}~\bibnamefont{Steane}},
  \bibinfo{journal}{Nature} \textbf{\bibinfo{volume}{399}},
  \bibinfo{pages}{124} (\bibinfo{year}{1999}).

\bibitem[{\citenamefont{Gottesman and Chuang}(1999)}]{gottesman:qc1999a}
\bibinfo{author}{\bibfnamefont{D.}~\bibnamefont{Gottesman}} \bibnamefont{and}
  \bibinfo{author}{\bibfnamefont{I.~L.} \bibnamefont{Chuang}},
  \bibinfo{journal}{Nature} \textbf{\bibinfo{volume}{402}},
  \bibinfo{pages}{390} (\bibinfo{year}{1999}).

\bibitem[{\citenamefont{Gottesman and Preskill}(1999)}]{gottesman:qc2000b}
\bibinfo{author}{\bibfnamefont{D.}~\bibnamefont{Gottesman}} \bibnamefont{and}
  \bibinfo{author}{\bibfnamefont{J.}~\bibnamefont{Preskill}}
  (\bibinfo{year}{1999}), \bibinfo{note}{unpublished analysis of the accuracy
  threshold.}

\bibitem[{\citenamefont{Knill et~al.}(2001)\citenamefont{Knill, Laflamme, and
  Milburn}}]{knill:qc2000e}
\bibinfo{author}{\bibfnamefont{E.}~\bibnamefont{Knill}},
  \bibinfo{author}{\bibfnamefont{R.}~\bibnamefont{Laflamme}}, \bibnamefont{and}
  \bibinfo{author}{\bibfnamefont{G.}~\bibnamefont{Milburn}},
  \bibinfo{journal}{Nature} \textbf{\bibinfo{volume}{409}}, \bibinfo{pages}{46}
  (\bibinfo{year}{2001}).

\bibitem[{\citenamefont{Panel}(2002)}]{hughes:qc2002a}
\bibinfo{author}{\bibfnamefont{T.~E.} \bibnamefont{Panel}},
  \bibinfo{type}{Tech. Rep.} \bibinfo{number}{LAUR-02-6900},
  \bibinfo{institution}{Los Alamos National Laboratory} (\bibinfo{year}{2002}),
  \bibinfo{note}{produced for ARDA}.

\bibitem[{\citenamefont{Aharonov}(2002)}]{aharonov:qc2002a}
\bibinfo{author}{\bibfnamefont{D.}~\bibnamefont{Aharonov}}
  (\bibinfo{year}{2002}), \bibinfo{note}{talk at QIP, IBM, Jan 16, 2002}.

\bibitem[{\citenamefont{Steane}(2003)}]{steane:qc2002a}
\bibinfo{author}{\bibfnamefont{A.~M.} \bibnamefont{Steane}},
  \bibinfo{journal}{Phys. Rev. A} \textbf{\bibinfo{volume}{68}},
  \bibinfo{pages}{042322/1} (\bibinfo{year}{2003}).

\bibitem[{\citenamefont{Steane}(2002)}]{steane:qc2002b}
\bibinfo{author}{\bibfnamefont{A.~M.} \bibnamefont{Steane}}
  (\bibinfo{year}{2002}), \bibinfo{note}{quant-ph/0202036}.

\bibitem[{\citenamefont{Fenner et~al.}(2003)\citenamefont{Fenner, Green, Homer,
  and Zhang}}]{fenner:qc2003a}
\bibinfo{author}{\bibfnamefont{S.}~\bibnamefont{Fenner}},
  \bibinfo{author}{\bibfnamefont{F.}~\bibnamefont{Green}},
  \bibinfo{author}{\bibfnamefont{S.}~\bibnamefont{Homer}}, \bibnamefont{and}
  \bibinfo{author}{\bibfnamefont{Y.}~\bibnamefont{Zhang}}
  (\bibinfo{year}{2003}), \bibinfo{note}{quant-ph/0312209}.

\bibitem[{\citenamefont{Calderbank et~al.}(1998)\citenamefont{Calderbank,
  Rains, Shor, and Sloane}}]{calderbank:qc1996b}
\bibinfo{author}{\bibfnamefont{A.~R.} \bibnamefont{Calderbank}},
  \bibinfo{author}{\bibfnamefont{E.~M.} \bibnamefont{Rains}},
  \bibinfo{author}{\bibfnamefont{P.~W.} \bibnamefont{Shor}}, \bibnamefont{and}
  \bibinfo{author}{\bibfnamefont{N.~J.~A.} \bibnamefont{Sloane}},
  \bibinfo{journal}{IEEE Trans. Inf. Theory} \textbf{\bibinfo{volume}{44}},
  \bibinfo{pages}{1369} (\bibinfo{year}{1998}).

\bibitem[{\citenamefont{Bernstein and Vazirani}(1997)}]{bernstein:qc1997a}
\bibinfo{author}{\bibfnamefont{E.}~\bibnamefont{Bernstein}} \bibnamefont{and}
  \bibinfo{author}{\bibfnamefont{U.}~\bibnamefont{Vazirani}},
  \bibinfo{journal}{SIAM J. Comput.} \textbf{\bibinfo{volume}{26}},
  \bibinfo{pages}{1411} (\bibinfo{year}{1997}).

\bibitem[{\citenamefont{MacWilliams and Sloane}(1977)}]{macwilliams:qc1977a}
\bibinfo{author}{\bibfnamefont{F.~J.} \bibnamefont{MacWilliams}}
  \bibnamefont{and} \bibinfo{author}{\bibfnamefont{N.~J.}
  \bibnamefont{Sloane}}, \emph{\bibinfo{title}{The Theory of Error-Correcting
  Codes}} (\bibinfo{publisher}{North-Holland Publishing Company},
  \bibinfo{year}{1977}).

\bibitem[{\citenamefont{Cleve and Gottesman}(1997)}]{cleve:qc1996b}
\bibinfo{author}{\bibfnamefont{R.}~\bibnamefont{Cleve}} \bibnamefont{and}
  \bibinfo{author}{\bibfnamefont{D.}~\bibnamefont{Gottesman}},
  \bibinfo{journal}{Phys. Rev. A} \textbf{\bibinfo{volume}{56}},
  \bibinfo{pages}{76} (\bibinfo{year}{1997}).

\bibitem[{\citenamefont{Grassl et~al.}(1997)\citenamefont{Grassl, Beth, and
  Pellizari}}]{grassl:1996a}
\bibinfo{author}{\bibfnamefont{M.}~\bibnamefont{Grassl}},
  \bibinfo{author}{\bibfnamefont{T.}~\bibnamefont{Beth}}, \bibnamefont{and}
  \bibinfo{author}{\bibfnamefont{T.}~\bibnamefont{Pellizari}},
  \bibinfo{journal}{Phys. Rev. A} \textbf{\bibinfo{volume}{56}},
  \bibinfo{pages}{33} (\bibinfo{year}{1997}).

\bibitem[{\citenamefont{Cleve et~al.}(1999)\citenamefont{Cleve, Gottesman, and
  Lo}}]{cleve:qc1999a}
\bibinfo{author}{\bibfnamefont{R.}~\bibnamefont{Cleve}},
  \bibinfo{author}{\bibfnamefont{D.}~\bibnamefont{Gottesman}},
  \bibnamefont{and} \bibinfo{author}{\bibfnamefont{H.-K.} \bibnamefont{Lo}},
  \bibinfo{journal}{Phys. Rev. Lett.} \textbf{\bibinfo{volume}{83}},
  \bibinfo{pages}{648} (\bibinfo{year}{1999}).

\end{thebibliography}

\end{document}